\documentclass[aip, jap, preprint, 11pt]{revtex4-1}

\usepackage{mathtools}
\usepackage[colorlinks=true,
    citecolor=blue,
    linkcolor=black,
    urlcolor=blue,
    pagebackref=false]{hyperref}
\usepackage{array}

\begin{document}

\title{Improved optical standing-wave beam splitters for
dilute
Bose--Einstein condensates} 

\author{Mary Clare Cassidy}
\affiliation{Department of Electrical Engineering and Computer Science, U.S. Military Academy, West Point NY 10996}

\author{Malcolm G. Boshier}
\affiliation{Materials Physics \& Applications Division, Los Alamos National Laboratory, Los Alamos NM 87545}

\author{Lee E. Harrell}
\email[]{lee.harrell@westpoint.edu}
\affiliation{Department of Physics and Nuclear Engineering, U.S. Military Academy, West Point NY 10996}

\date{\today}

\begin{abstract}
Bose--Einstein condensate (BEC)-based atom interferometry exploits low temperatures and long coherence lengths to facilitate high-precision measurements. Progress in atom interferometry promises improvements in navigational devices like gyroscopes and accelerometers, as well as applications in fundamental physics such as accurate determination of physical constants. Previous work demonstrates that beam splitters and mirrors for coherent manipulation of dilute BEC momentum in atom interferometers can be implemented with sequences of non-resonant standing-wave light pulses. While previous work focuses on the optimization of the optical pulses' amplitude and duration to produce high-order momentum states with high fidelity, we explore how varying the shape of the optical pulses affects optimal beam-splitter performance, as well as the effect of pulse shape on the sensitivity of optimized parameters in achieving high fidelity in high-momentum states.
In simulations of two-pulse beam splitters utilizing optimized square, triangle, and sinc-squared pulse shapes applied to dilute BECs,
we, in some cases, reduce parameter sensitivity by an order of magnitude while maintaining fidelity. 
\end{abstract}

\maketitle 

\section{Introduction}\label{sec:introduction}
Atom interferometry exploits the properties of matter waves in the same way that optical interferometry exploits the properties of electromagnetic waves.\cite{RevModPhys.81.1051,wang2005atom, garcia2006bose, burke2008confinement, bongs2019taking} 
In place of the beam splitters and mirrors used in optical interferometry, atom interferometry uses optical standing waves to split and recombine matter waves. \cite{wu2005splitting, stickney2008theoretical, edwards2010momentum} 
Splitting and recombining matter waves creates path-dependent phase differences that result in interference patterns. 
These patterns are readily coupled to the surrounding environment through the sensitivity of atomic quantum phases to electromagnetic fields and other local effects.
This sensitivity to electromagnetic fields gives atom interferometry an advantage over optical interferometry in high-precision measurement tools and sensor applications because optical interferometry requires a secondary transduction to measure these
fields. \cite{RevModPhys.81.1051} Atom interferometers also have increased sensitivity to inertial forces and thus advance technologies like gyroscopes and navigational devices, particularly in applications of oceanic and space exploration.\cite{bongs2019taking}

Thermal vapors,\cite{crookston2005microchip, bongs2019taking} stationary Bose--Einstein condensates (BECs),\cite{wang2005atom, garcia2006bose, crookston2005microchip} and atom lasers \cite{borde1995amplification, RevModPhys.81.1051} are all potential atom sources for atom interferometry. 
BECs are of particular interest as atom sources because they are confined in both momentum and coordinate space, giving rise to the long coherence lengths necessary for precision interferometry.\cite{dalfovo1999theory, jamison2011atomic} Because the atoms of a BEC are in a collective state of indistinguishable particles, they can be considered a single matter wave and described by a single wave function.\cite{dalfovo1999theory}

\begin{figure}
  \includegraphics[width=3.37in]{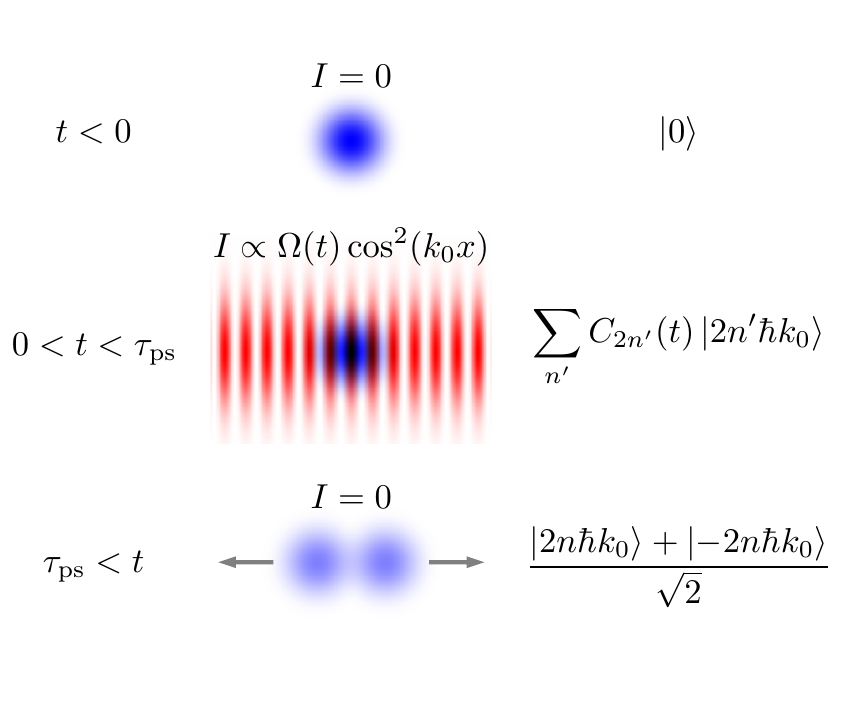}
  \caption{\label{Wave_Fig}Operating concept of the optical standing-wave beam splitter. Prior to the optical pulse sequence, the BEC is in a zero-momentum state (top row). Between times $t=0$ and $\tau_\mathrm{ps}$, the BEC is subjected to a sequence of optical pulses during which it is in a time-dependent linear superposition of states with momentum in integer multiples of $2\hbar k_0$ (middle row). Following the optical pulse sequence, the BEC is left in the target beam-splitter state of definite momentum (bottom row).}
\end{figure}

{Kapitza--Dirac optical pulses are used to create beam splitters for the matter waves used in atom interferometry.\cite{gould1986diffraction,wu2005splitting, stickney2008theoretical, edwards2010momentum} These non-resonant optical standing-wave pulses} excite the stationary BEC into a linear superposition of states with linear momentum $\pm2n\hbar k_0$, where $k_0$ is the optical wave number and $n$ is a positive integer.\cite{wu2005splitting}
Optimal protocols for splitting BECs produce states of definite $n$ with high fidelity  as illustrated in Fig.~\ref{Wave_Fig}.

 A critical element of the atom interferometer is the matter-wave beam splitter. Beam splitters that are implemented with light-pulse sequences producing states of high $n$ while maintaining fidelity  are advantageous
 because of the increased sensitivity associated with the shorter wavelengths and increased velocities associated with these states.
\cite{xiong2011manipulating, wu2005splitting} Faster velocities enable split matter waves to travel further before recombination, thus increasing the coupling to the environment and the resulting signal.

Previous work has looked at optimizing the splitting of 
dilute
BECs using various square pulse sequences to improve the  post-splitting fidelity of high-momentum states.\cite{wu2005splitting, edwards2010momentum, xiong2011manipulating, wu2007demonstration, hughes2007high} 
As shown in Refs.~\onlinecite{wu2005splitting} and \onlinecite{xiong2011manipulating}, two-pulse equal-amplitude square pulse sequences are highly effective at producing populations in high-order momentum states; however better fidelity can be achieved with with sequences of three or more pulses.\cite{edwards2010momentum} Additionally, M\"uller \emph{et al}. have explored optical pulses with Gaussian envelopes in the context of matter-wave Bragg mirrors.\cite{muller2008atom}

In this paper, we explore how deviating from the more commonly studied equal-amplitude square optical pulses and considering two-pulse sequences with other envelopes affects the fidelity of high-order momentum target states and the sensitivity of beam-splitter performance to experimental variations from optimal pulse parameters. Sensitivity to variations in optimized parameters is a critical consideration in determining whether a particular pulse-parameter tuning for a high-fidelity splitting can be implemented in practice. 
We study square pulses of unequal amplitudes, triangular-shaped pulses, and sinc-squared shaped pulses, although our methods are applicable to arbitrary pulse envelopes. 

As explained in Ref.\ \onlinecite{edwards2010momentum}, for sufficiently short and intense optical pulses (the Raman-Nath regime), the effect on the BEC wave function is completely determined by the area of its envelope and, therefore, insensitive to its shape. Mathematically, this conclusion is a consequence of neglecting the evolution of the BEC arising from the kinetic energy term in the Hamiltonian. In order to account for the effect of envelope shape we extend the Raman-Nath analysis into the quasi-Bragg regime\cite{muller2008atom,gadway2009analysis} by retaining the kinetic energy Hamiltonian and considering pulses that violate the strict Raman-Nath condition.

The remainder of the paper is organized as follows: in Sec.~\ref{sec:method}, we describe a numerical method to extend the analysis presented in Ref.~\onlinecite{wu2005splitting}. Sec.~\ref{sec:results} presents the results for optimized pulse parameters for $\pm2nk_0\hbar$ target states where $n$ varies from $n=1$ to $n=4$ for three pulse shapes. Additionally, we present the results of convergence tests of our numerical approximations and analysis of the sensitivity of target state fidelity  to variations in pulse parameters from optimal tuning. In Sec.~\ref{sec:discussion}, we analyze the results and discuss the implications for considering alternatives to square pulses. Section~\ref{sec:conclusion} summarizes our findings and discusses potential future work. 

\section{Method}\label{sec:method}
The present work is motivated in the applied context of experimental atom interferometry implemented in dilute BECs. However, in the dilute, non-interacting limit, preparing the atoms from a BEC is not essential to the result. Other techniques such as delta kick cooling can achieve atom ensembles with the necessary spatial localization and narrow momentum distribution about zero.\cite{ammann1997delta} For economy of language we refer to such ensembles as dilute BECs, but the reader should be mindful of the broader context in which the results apply. We return to the question of the necessary momentum distribution in Sec.\ \ref{sec:discussion}. 

We seek to model the behavior of an initially stationary
dilute
BEC subject to a coherent optical standing wave of intensity
\begin{align}
I(x,t) &= I_0(t)\cos^2(k_0x)\nonumber\\
       &= I_0(t)\frac{1+\cos(2k_ox)}{2}, 
\label{eq:intensity}
\end{align}
and wave number $k_0$ by solving the one-dimensional single-atom Schr{\"o}dinger equation
\begin{equation}
i\dot\psi(x,t)=\left(-\frac{\hbar}{2m}\frac{\,\partial\,^2}{\partial x^2}+\Omega(t)\cos(2k_0x)\right)\psi(x,t)
\label{eq:one_atom_SE}
\end{equation}
in the manner presented by Wu \emph{et al}.\cite{wu2005splitting}
The term $\Omega(t)\cos(2k_0x)$ is the light shift potential, which arises from the interaction of the optical standing wave with a near-resonant electronic transition in the BECs atoms via the ac Stark effect.\cite{meystre2001atom} $\Omega(t)$ is the strength of the coupling, which is proportional to the instantaneous intensity $I(t)$ of the optical standing wave.
Note that in going from Eq.\ (\ref{eq:intensity}) to Eq.\ (\ref{eq:one_atom_SE}) we drop the spatially constant part of the light shift potential as it only contributes an uninteresting overall phase factor to the wave function.

Writing the wave function in the general form
\begin{multline}
\psi(x,t)=
\int^{k_0}_{-k_0}dk\,e^{ikx}\left[ C_0(k,t) \vphantom{\frac12} \right. \\
\left.  + \sum^{\infty}_{n'=1}\left(C^+_{2n'}(k,t)
\frac{e^{i2n'k_0x}+e^{-i2n'k_0x}}{\sqrt{2}}
+ C^-_{2n'}(k,t)
\frac{e^{i2n'k_0x}-e^{-i2n'k_0x}}{\sqrt{2}}
\right)\right],
\label{eq:plus_minus_basis}
\end{multline}
Eq.\ (\ref{eq:one_atom_SE}) becomes a system of equations for the evolution of the coefficients $C_0(k,t)$, $C^+_{2n}(k,t)$, and $C^-_{2n}(k,t)$ for all $n$. These coefficients are the amplitudes of the beam-splitter states of the BEC. Because the amplitudes corresponding to different values of $k$ evolve independently under Eq.\ (\ref{eq:one_atom_SE}) and the BEC is taken to be initially at rest, we consider only the amplitudes corresponding to $k=0$ and write them as $C_0$, $C^+_{2n}$, and $C^-_{2n}$, dropping explicit reference to their $k$ dependence. This approximation is equivalent to ignoring the finite extent of the BEC relative to the wavelength of the optical standing wave. So restricted, the beam-splitter amplitudes obey
\begin{align}
\dot{C}_0&=-i\omega_\mathrm{r}\sqrt{2}\frac{\Omega(t)}{2\omega_\mathrm{r}}C_{2}^+,\label{eq:C_0}\\
\dot{C}_2^+ &=-i\omega_\mathrm{r}\left(4C_2^++\sqrt{2}\frac{\Omega(t)}{2\omega_\mathrm{r}}C_0+\frac{\Omega(t)}{2\omega_\mathrm{r}}C_4^+\right),\label{eq:C_+_2}\\
\dot{C}_2^- &=-i\omega_\mathrm{r}\left(4C_2^-+\frac{\Omega(t)}{2\omega_\mathrm{r}}C_4^-\right),
\label{eq:C_-_2}
\end{align}
and, for $n>1$,
\begin{align}
\dot{C}_{2n}^+ &= -i\omega_\mathrm{r}\left(4n^2C_{2n}^+
+\frac{\Omega(t)}{2\omega_\mathrm{r}}\left(C_{2(n+1)}^++C_{2(n-1)}^+\right)\right),
\label{eq:C_+_2n}\\
\dot{C}_{2n}^- &=-i\omega_\mathrm{r}\left(4n^2C_{2n}^-+\frac{\Omega(t)}{2\omega_\mathrm{r}}\left(C_{2(n+1)}^-+C_{2(n-1)}^-\right)\right),
\label{eq:C_-_2n}
\end{align}
where $\hbar\omega_\mathrm{r}=\hbar^2 k_0^2\,/\,2m$ is the photon recoil energy of an atom with mass $m$.

For a stationary BEC, the initial state is $C_0=1$ and $C^{\pm}_{2n}=0$ for all $n>0$. Noting that, for $k=0$, the $C^-_{2n}$ are completely decoupled from $C_0$ and the $C^+_{2n}$, we have immediately that $C^-_{2n}(t)=0$ for all $t$, and Eqs. (\ref{eq:C_-_2}) and (\ref{eq:C_-_2n}) can be ignored. Further, the rate at which the $C^+_{2n}$ states are populated decreases with increasing $n$ and is negligible\cite{wu2005splitting} for $n \gg \sqrt{\Omega(t)\,/\,8\omega_\mathrm{r}}$. Accordingly, we truncate the representation of the BEC state by taking $C^+_{2n}=0$ for $n$ greater than some appropriately chosen value $N$. In this approximation, Eqs.\ (\ref{eq:C_0}), (\ref{eq:C_+_2}), and (\ref{eq:C_+_2n}) can be written in the form
\begin{equation}
\dot{C}^{+}=-i\omega_\mathrm{r}A(t)C^{+},
\label{eq:eom_matrix}
\end{equation} 
where $C^+ = [C_0,C_2^+,C_4^+, \ldots ,C_{2N}]^\mathrm{T}$ and $A(t)$ is the real, symmetric, $(N+1)\times (N+1)$ matrix
\begin{equation}
A(t)=
\begin{bmatrix}
0      & 0      & 0      & \cdots & 0\\
0      & 4      & 0      & \cdots & 0\\
0      & 0      & 16     & \cdots & 0\\
\vdots & \vdots & \vdots & \ddots & \vdots\\
0      & 0      & 0      & \cdots & (2N)^2
\end{bmatrix}
+\frac{\Omega(t)}{2\omega_\mathrm{r}}
\begin{bmatrix}
0        & \sqrt{2} & 0      & 0      & \cdots & 0 & 0\\
\sqrt{2} & 0        & 1      & 0      & \cdots & 0 & 0\\
0        & 1        & 0      & 1      & \cdots & 0 & 0\\
0        & 0        & 1      & 0      & \cdots & 0 & 0\\
\vdots   & \vdots   & \vdots & \vdots & \ddots & \vdots &\vdots\\
0        & 0        & 0      & 0      & \cdots & 0 & 1\\
0        & 0        & 0      & 0      & \cdots & 1 & 0
\end{bmatrix},
\label{eq:A_matrix}
\end{equation}
which depends only on the unitless ratio $\Omega(t)/\omega_\mathrm{r}$.

For constant $\Omega$, the formal solution to Eq. (\ref{eq:eom_matrix}) is
\begin{equation}
C^+(t)=e^{-i\omega_\mathrm{r}tA}C^+(0).
\label{eq:matrix_solution}
\end{equation}
The matrix exponential $e^{-i\omega_\mathrm{r}(t-t_0)A}$ in Eq.\ (\ref{eq:matrix_solution}) can be evaluated by writing $A=UDU^\mathrm{T}$, where the matrix $U$ is unitary, the matrix $D$ is diagonal, and both are real. This decomposition is always possible by choosing the diagonal elements of $D$ to be the eigenvalues of $A$ and the columns of $U$ to be corresponding orthonormal eigenvectors.  The solution is then
\begin{equation}
C^+(t)=U
\begin{bmatrix}
e^{-i\omega_\mathrm{r}tD_{0}}   & \cdots & 0\\
\vdots                                & \ddots & \vdots \\
0                                     & \cdots & e^{-i\omega_\mathrm{r}tD_{N}}
\end{bmatrix}U^\mathrm{T}C^+(0).
\label{eq:decomposed_matrix_solution}
\end{equation}

\subsection{Square Pulses}
In Ref.~\onlinecite{wu2005splitting}, Wu \emph{et al}. propose selectively exciting BEC beam-splitter states using two equal-intensity square optical pulses separated by a period of unperturbed evolution. We generalize this protocol by allowing the optical pulses to be of differing intensities such that
\begin{equation}
\Omega(t)=
\begin{cases}
\Omega_1, &0 \leq t \leq \tau_1,\\
0, & \tau_1 < t < \tau_1+\tau_2,\\
\Omega_2, &\tau_1+\tau_2 \leq t \leq \tau_\mathrm{ps},
\end{cases}
\label{eq:Omega_square}    
\end{equation}
where $\Omega_{1,2}$ and $\tau_{1,3}$ are the strength and durations of the two optical pulses, $\tau_2$ is the time between the pulses, and $\tau_\mathrm{ps}=\tau_1+\tau_2+\tau_3$ is the total duration of the pulse sequence.

To find the state $C^+_f$ of the BEC  following this square-pulse sequence, we apply Eq.\ (\ref{eq:matrix_solution}) three times,
\begin{equation}
C^+_f=e^{-i\omega_r\tau_3A_3}e^{-i\omega_r\tau_2A_2}e^{-i\omega_r\tau_1A_1}C^+(0),
\label{eq:solve_two_square_pulses}
\end{equation}
to evolve the state vector through each of the constant-$\Omega$ periods. The decompositions of the tri-diagonal matrices $A_i$, which must be done independently for each value of $\Omega(t)$, are carried out numerically. Otherwise, Eq.\ (\ref{eq:solve_two_square_pulses}) is an exact solution of Eq.\ (\ref{eq:eom_matrix}). 

\subsection{Shaped Pulses}
In addition to square pulses, we study the selective excitation of beam-splitter states using optical pulses with triangle and sinc-squared envelopes. While  representing  a  limited  sample, these two shapes exhibit distinct qualitative differences from  square  pulses. Unlike  the  square  pulses, triangle and sinc-squared pulse shapes are continuous, with the sinc-squared  shape also having a continuous first derivative.

The profiles of these shaped-pulse sequences are defined by the same five parameters as the square-pulse sequence and are given by
\begin{equation}
\Omega(t)=
\begin{cases}
\Omega_1\frac{2t}{\tau_1} &0 \leq t \leq \frac{\tau_1}{2}\\
\Omega_1\frac{2(\tau_1-t)}{\tau_1} &\frac{\tau_1}{2} \leq t \leq \tau_1\\
0 & \tau_1 \leq t \leq \tau_1+\tau_2\\
\Omega_2\frac{2(t-\tau_1-\tau_2)}{\tau_3} &\tau_1+\tau_2 \leq t \leq \tau_\mathrm{ps}-\frac{\tau_3}{2}\\
\Omega_2\frac{2(\tau_\mathrm{ps}-t)}{\tau_3} &\tau_\mathrm{ps}-\frac{\tau_3}{2} \leq t \leq \tau_\mathrm{ps},
\end{cases}
\label{eq:Omega_triangle}    
\end{equation}
and
\begin{equation}
\Omega(t)=
\begin{cases}
\Omega_1\operatorname{sinc}^2\left( \frac{2t-\tau_1}{\tau_1}\right) &0 \leq t \leq \tau_1\\
0 & \tau_1 < t < \tau_1+\tau_2\\
\Omega_2\operatorname{sinc}^2\left( \frac{2(t-\tau_\mathrm{ps})+\tau_3}{\tau_3}\right) &\tau_1+\tau_2 \leq t \leq \tau_\mathrm{ps},
\end{cases}
\label{eq:Omega_sinc_sqrd}    
\end{equation}
where we have used the normalized $\operatorname{sinc}(x)=\sin(\pi x)/\pi x$ function.

The methods that are described above for calculating the final state of the BEC after a sequence of square pulses cannot be applied directly to the triangle or sinc-squared pulse sequences because $\Omega(t)$ is not piece-wise constant. However, $\Omega(t)$ can be approximated by slicing a continuously varying pulse into a series of $N_\mathrm{s}$ square pulses, allowing the final state to be calculated by repeated application of Eq.\ (\ref{eq:matrix_solution}), once for each slice. Adopting the notation $A_{1\,k}$ for the matrix $A(\Omega(t))$ evaluated at $t=\tau_1(k-1/2)/N_\mathrm{s}$ for $k=1,2,...,N_\mathrm{s}$, and similarly for the second pulse, the approximate post-optical-pulse-sequence state is
\begin{equation}
C^+_f=
\left(\,\prod_{k=1}^{N_\mathrm{s}}e^{-i\frac{\omega_r\tau_3}{N_\mathrm{s}}A_{3\,k}}\right)
e^{-i\omega_r\tau_2A_2}
\left(\,\prod_{k=1}^{N_\mathrm{s}}e^{-i\frac{\omega_r\tau_1}{N_\mathrm{s}}A_{1\,k}}\right)C^+(0).
\label{eq:sliced_evolution}
\end{equation}
The calculated final state of the BEC converges toward the exact result as $N_\mathrm{s}$ increases. As is detailed in Sec.\ \ref{sec:results}, we find satisfactory convergence using 50 slices for each pulse for the parameter ranges that we consider.

{The method described above is implemented in Python using the matrix algebra functionalities of the numpy,\cite{harris2020array} scipy.linalg, and scipy.sparse packages.\cite{2020SciPy-NMeth}
On the basis of informal testing, we find this method to be an order of magnitude or more faster than integrating Eqs.\ (\ref{eq:C_0})--(\ref{eq:C_-_2n})
using the numerical ODE integrators scipy.integrate.odeint (LSODA method) and scipy.integrate.sole-ivp (BDF method).\cite{2020SciPy-NMeth, osti_145724,byrne1975polyalgorithm}}

\subsection{Experimental Parameter Optimization}
We seek a set of pulse parameters for selective excitation of the $\pm2n\hbar k_0$ beam-splitter state by maximizing the post-pulse-sequence fidelity $|C_{f\,2n}^+|^2$ with respect to variation in $\Omega_1$, $\Omega_2$, $\tau_1$ $\tau_2$, and $\tau_3$. 
To this end, we employ the Sequential Least Squares Programming (SLSQP) algorithm as implemented in the scientific computation library SciPy.\cite{2020SciPy-NMeth} After obtaining a set of optimized parameters $(\Omega_1^*, \Omega_2^*,\tau_1^*, \tau_2^*, \tau_3^*)$, we plot $|C_{f\,2n}^+|^2$ as a function of the fractional deviation of each parameter 
to ensure that the algorithm has returned a maximum. The curvature of the plots provides a quantitative measure of the sensitivity of the result's fidelity to variations in the experimental parameters. 


\section{Results}\label{sec:results}

\subsection{Pulse Parameter Optimization}

We apply the methods described in Sec.\ \ref{sec:method}, with $N_\mathrm{s}=50$ slices and $N=24$ coefficients, to identify optimal pulse parameters for high fidelity of target states $n=1\mathrm{\ through \ } n=4$. The results of our optimizations for square, sinc-squared, and triangle pulse shapes are presented in Tabs.\ \ref{table:equal-amp-square-pulses}--\ref{table:triangular-pulses}. The results for equal-amplitude square pulses, Tab.\ \ref{table:equal-amp-square-pulses}, show good agreement with comparable calculations presented in Ref.\ \onlinecite{wu2005splitting}.

\begin{table}
\caption{\label{table:equal-amp-square-pulses}Optimal pulse parameters for equal-amplitude square light pulses. Units and notation are chosen to facilitate direct comparison with Ref.~\onlinecite{wu2005splitting}.}
\begin{tabular}{>{\centering}p{0.55in}
                >{\centering}p{0.35in}
								>{\centering}p{0.05in}
								>{\centering}p{0.35in}
								>{\centering}p{0.5in}
								>{\centering}p{0.5in}
								>{\centering\arraybackslash}p{0.5in}}
\hline
\multicolumn{2}{c}{Target State} & ~~ &%
\multicolumn{4}{c}{Optimal Parameters} \\[0.06in]
$\displaystyle \pm2n\hbar k_0$  &%
$\displaystyle |C^+_{2n}|^2$ &%
&%
$\displaystyle \frac{\Omega}{\omega_r}$ &%
$\displaystyle \frac{\omega_r\tau_1}{2\pi}$ &%
$\displaystyle \frac{\omega_r\tau_2}{2\pi}$ &%
$\displaystyle \frac{\omega_r\tau_3}{2\pi}$ \\[0.08in] 
\cline{1-2}\cline{4-7} \\[-0.1in]
$\pm2\hbar k_0$   & .999   & & 2.81       & 0.0841      & 0.142      & 0.0795  \\
$\pm4\hbar k_0$   & .991   & & 13.3       & 0.147       & 0.113      & 0.171   \\
$\pm6\hbar k_0$   & .968   & & 34.6       & 0.0289      & 0.0882     & 0.0302  \\
$\pm8\hbar k_0$   & .923   & & 58.3       & 0.0662      & 0.0455     & 0.0237  \\   
\hline
\end{tabular}
\end{table}

\begin{table}
\caption{\label{table:unequal-amp-square-pulses} Optimal pulse parameters for unequal-amplitude square light pulses. There are three optimal parameter tunings provided for the $\pm8\hbar k_0$ state, demonstrating that multiple local maximums for target-state fidelity are common. However, we find that higher fidelity may come at the cost of significantly higher sensitivity to variations in the pulse parameters. The trade-off between fidelity and parameter sensitivity is discussed in Sec.~\ref{sec:discussion}.}

\begin{tabular}{>{\centering}p{0.55in}
                >{\centering}p{0.35in}
								>{\centering}p{0.05in}
								>{\centering}p{0.35in}
								>{\centering}p{0.4in}
								>{\centering}p{0.5in}
								>{\centering}p{0.5in}
								>{\centering\arraybackslash}p{0.5in}}
\hline
\multicolumn{2}{c}{Target State} & ~~ &%
\multicolumn{5}{c}{Optimal Parameters} \\[0.06in]
$\displaystyle \pm2n\hbar k_0$  &%
$\displaystyle |C^+_{2n}|^2$ &%
&%
$\displaystyle \frac{\Omega_1}{\omega_r}$ &%
$\displaystyle \frac{\Omega_2}{\omega_r}$ &%
$\displaystyle \frac{\omega_r\tau_1}{2\pi}$ &%
$\displaystyle \frac{\omega_r\tau_2}{2\pi}$ &%
$\displaystyle \frac{\omega_r\tau_3}{2\pi}$ \\[0.08in] 
\cline{1-2}\cline{4-8} \\[-0.1in]
$\pm2\hbar k_0$   & .999   & & 2.83          & 2.80        & 0.102         & 0.107        & 0.0971 \\
$\pm4\hbar k_0$   & .996   & & 15.1          & 13.3        & 0.136         & 0.113        & 0.174  \\
$\pm6\hbar k_0$   & .975   & & 32.3          & 34.7        & 0.0303        & 0.0879       & 0.0305 \\
$\pm8\hbar k_0$   & .903   & & 61.1          & 48.9        & 0.0623        & 0.0421       & 0.0667 \\
$\pm8\hbar k_0$   & .943   & & 63.5          & 53.9        & 0.0622        & 0.0454       & 0.0241 \\
$\pm8\hbar k_0$   & .982   & & 85.1          & 54.4        & 0.193         & 0.204        & 0.186  \\
\hline
\end{tabular}
\end{table}

\begin{table}
\caption{\label{table:sinc-sqaured-pulses}Optimal pulse parameters for sinc-squared light pulses.}
\begin{tabular}{>{\centering}p{0.55in}
                >{\centering}p{0.35in}
								>{\centering}p{0.05in}
								>{\centering}p{0.35in}
								>{\centering}p{0.4in}
								>{\centering}p{0.5in}
								>{\centering}p{0.5in}
								>{\centering\arraybackslash}p{0.5in}}
\hline
\multicolumn{2}{c}{Target State} & ~~ &%
\multicolumn{5}{c}{Optimal Parameters} \\[0.06in]
$\displaystyle \pm2n\hbar k_0$  &%
$\displaystyle |C^+_{2n}|^2$ &%
&%
$\displaystyle \frac{\Omega_1}{\omega_r}$ &%
$\displaystyle \frac{\Omega_2}{\omega_r}$ &%
$\displaystyle \frac{\omega_r\tau_1}{2\pi}$ &%
$\displaystyle \frac{\omega_r\tau_2}{2\pi}$ &%
$\displaystyle \frac{\omega_r\tau_3}{2\pi}$ \\[0.08in] 
\cline{1-2}\cline{4-8} \\[-0.1in]
$\pm2\hbar k_0$   & .999  & & 3.79           & 4.19          & 0.102       & 0.0822       & 0.167  \\
$\pm4\hbar k_0$   & .958  & & 17.7           & 16.9          & 0.169       & 0.0331       & 0.102  \\
$\pm6\hbar k_0$   & .866  & & 32.0           & 39.3          & 0.167       & 0.0710       & 0.0756 \\
$\pm8\hbar k_0$   & .958  & & 71.3           & 69.0          & 0.0426      & 0.0764       & 0.0494 \\   
\hline
\end{tabular}
\end{table}

\begin{table}
\caption{\label{table:triangular-pulses}Optimal pulse parameters for triangle light pulses.}

\begin{tabular}{>{\centering}p{0.55in}
                >{\centering}p{0.35in}
								>{\centering}p{0.05in}
								>{\centering}p{0.35in}
								>{\centering}p{0.4in}
								>{\centering}p{0.5in}
								>{\centering}p{0.5in}
								>{\centering\arraybackslash}p{0.5in}}
\hline
\multicolumn{2}{c}{Target State} & ~~ &%
\multicolumn{5}{c}{Optimal Parameters} \\[0.06in]
$\displaystyle \pm2n\hbar k_0$  &%
$\displaystyle |C^+_{2n}|^2$ &%
&%
$\displaystyle \frac{\Omega_1}{\omega_r}$ &%
$\displaystyle \frac{\Omega_2}{\omega_r}$ &%
$\displaystyle \frac{\omega_r\tau_1}{2\pi}$ &%
$\displaystyle \frac{\omega_r\tau_2}{2\pi}$ &%
$\displaystyle \frac{\omega_r\tau_3}{2\pi}$ \\[0.08in] 
\cline{1-2}\cline{4-8} \\[-0.1in]
$\pm2\hbar k_0$   & .999   & & 3.89        & 3.70         & 0.122          & 0.0878       & 0.139  \\
$\pm4\hbar k_0$   & .993   & & 23.3        & 27.2         & 0.0519         & 0.0687       & 0.0886 \\
$\pm6\hbar k_0$   & .971   & & 47.3        & 49.7         & 0.0439         & 0.0732       & 0.0453 \\
$\pm8\hbar k_0$   & .963   & & 79.2        & 76.8         & 0.0348         & 0.0849       & 0.0400 \\   
\hline
\end{tabular}
\end{table}

\subsection{Convergence Test}
The number of slices $N_\mathrm{s}$ and the number of coefficients $N$ used in our calculations affect the accuracy of the results for both the optimal parameters and the target state fidelity after the pulse sequence. To ensure accuracy, we conduct convergence tests and select $N_\mathrm{s}$ and $N$ for convergence to better than one part in $10^{-4}$. 

\begin{figure}
  \includegraphics[width=3.37in]{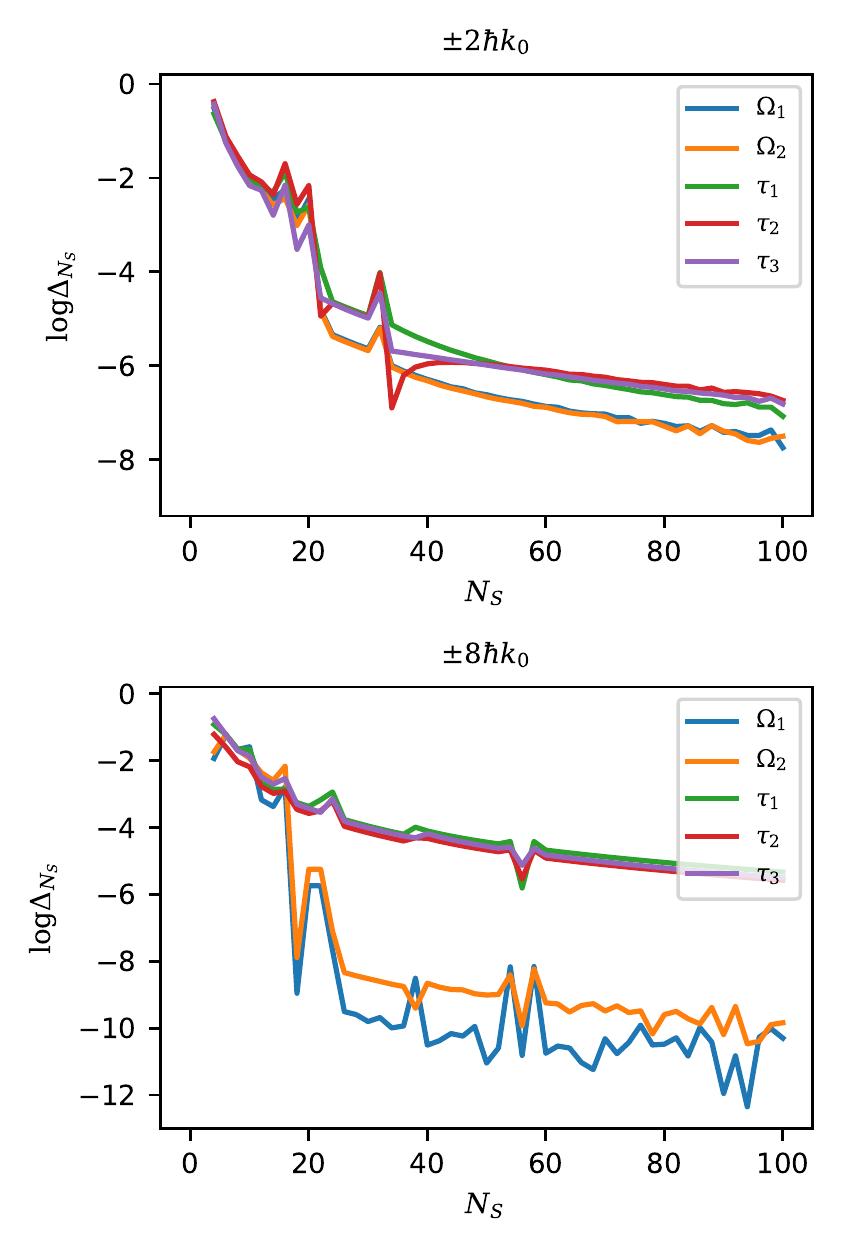}
  \caption{Convergence of optimal triangle pulse parameters with the number of slices. The estimated optimal pulse parameters change with the number of time slices $N_\mathrm{s}$ used to approximate the pulse shape, converging toward exact values as $N_\mathrm{s}$ increases. For each estimated optimal pulse parameter $i^* \in \{\Omega_1^*, \Omega_2^*, \tau_1^*, \tau_2^*, \tau_3^*\}$, we test convergence by calculating $\Delta_{N_\mathrm{s}}=\left|i^*(N_\mathrm{s}) - i^*(N_\mathrm{s} -2)\right|/i^*(N_\mathrm{s})$ as $N_\mathrm{s}$ varies in steps of 2 and plotting the results on a $\log_{10}$ scale. Plots for target states $\pm2\hbar k_0$ (top) and $\pm8\hbar k_0$ (bottom) are shown. Generally, convergence to one part in $10^{-4}$ or better is found for $N_\mathrm{s} > 50$.} 
  \label{slice-converge-param}
\end{figure}

Figure~\ref{slice-converge-param} shows the convergence of optimal optical-pulse parameters with increasing $N_\mathrm{s}$ for
the triangle pulse sequences. Regardless of target momentum state, we find convergence of optimal optical-pulse parameter values within $N_\mathrm{s}\ge50$ slices.
Notably, for higher-momentum target states, the pulse parameters corresponding to optical intensity are substantially less sensitive to $N_\mathrm{s}$ than are the parameters corresponding to pulse timing.
The convergence of target-state fidelity with increasing $N_s$ is found to be faster with all states converging by $N_\mathrm{s}\ge25$. Similar results are obtained for other pulse shapes.

Likewise, we check the convergence of optimal pulse parameters and target-state fidelity with increasing number of retained coefficients $N$. Truncation of the summation in Eq.~\ref{eq:plus_minus_basis} to $n' \le N$ ignores high-momentum beam-splitter states under the assumption that they have negligible population. As expected, convergence of optimal parameters and target-state fidelity requires proportionally more retained coefficients for higher-momentum target states. For states with momentum up to $\pm8\hbar k_0$, the optimal parameters converge to better than one part in $10^{-8}$ with $N\ge12$. Target state fidelity follows similar trends, converging to one part in $10^{-12}$ with $N\ge14$.
 
\subsection{Sensitivity}

\begin{figure*}
\includegraphics[width=6.69in]{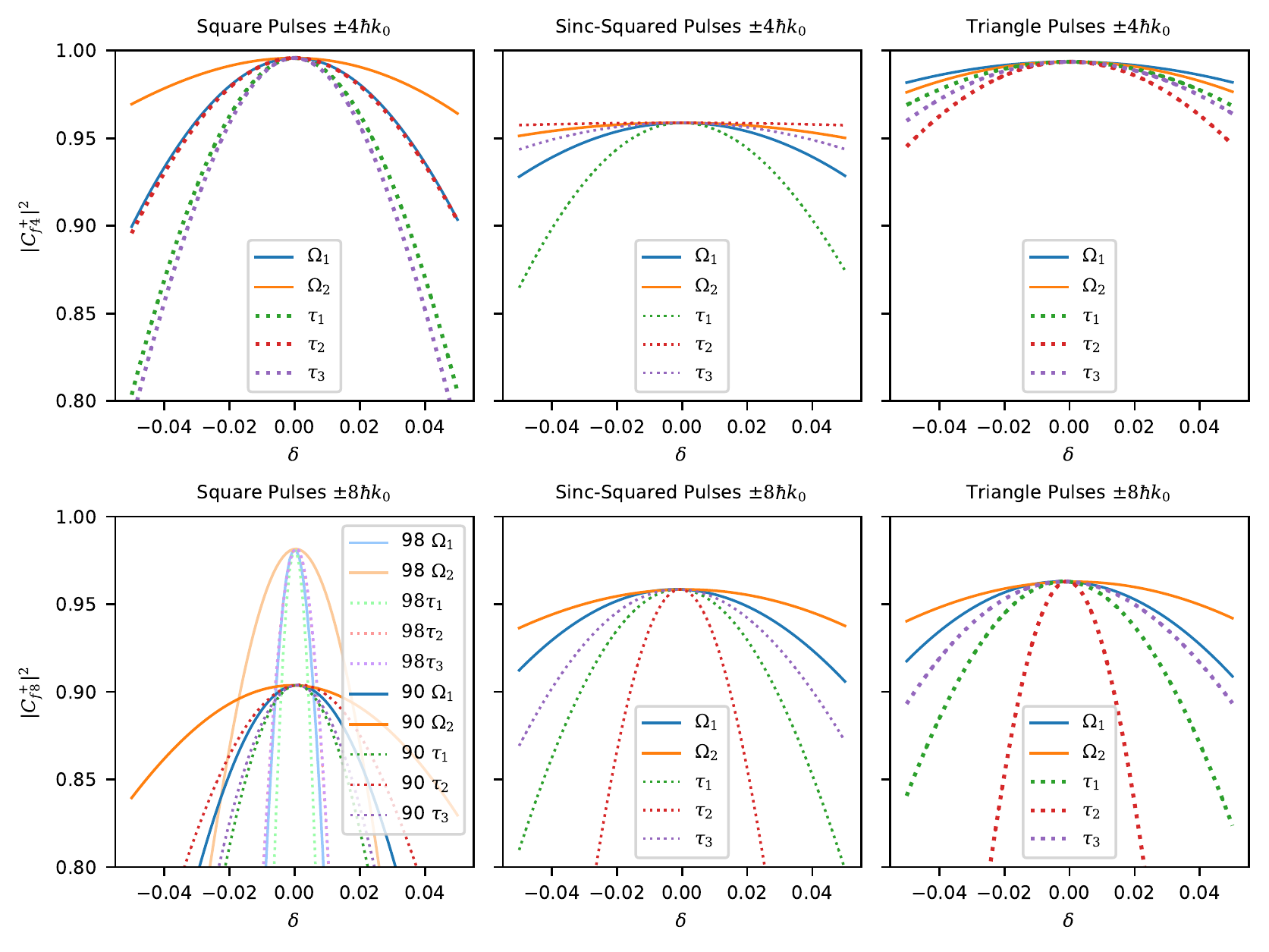}
\caption{\label{Sensitivity_pop_to_varied_param}
Sensitivity of fidelity to variations in the pulse parameters. The curvatures of these plots at $\delta=0$ provide a quantitative measure of the sensitivity of the optimization. For the $\pm4\hbar k$ and $\pm8\hbar k$ target states, we plot $|C_{f\,2n}^+|^2$ as a function of the fractional deviation $\delta(i)=(i-i^*)/i^*$ of each parameter $i \in \{\Omega_1, \Omega_2, \tau_1, \tau_2, \tau_3\}$ from its optimal value $i^*$, holding other parameters constant.}
\end{figure*}

In addition to maximizing the fidelity, we consider the sensitivity of the fidelity to deviations from the optimal pulse-parameter values. Figure \ref{Sensitivity_pop_to_varied_param} shows fidelity changes in the $\pm4\hbar k$ and $\pm8\hbar k$ states as pulse parameter values
$i \in \{\Omega_1, \Omega_2, \tau_1, \tau_2, \tau_3\}$
deviate from their optimum values $i^* \in \{\Omega_1^*, \Omega_2^*, \tau_1^*, \tau_2^*, \tau_3^*\}$.
For meaningful comparisons of parameters with different units, we plot the fidelity of the target states as a function of the relative variation of the pulse parameters
\begin{equation}
\delta(i)=(i-i^*)/i^*
\end{equation}
about their optimal values.
We then define sensitivity $S_i$ as the absolute value of the curvature of the plots in Fig.~\ref{Sensitivity_pop_to_varied_param} evaluated at zero-offset from the optimal parameter so that the change in the fidelity from its optimized value is
\begin{equation}
\Delta |C_{f\,2n}^+|^2  = -\frac{1}{2} S_i \delta^2(i).    
\end{equation}
Sensitivity values for various pulse shapes and tunings are reported in Tables \ref{table:sensitivity-square}--\ref{table:sensitivity-triangle}.

The sensitivities of the pulse parameter tunings are compared on the basis of the most sensitive amplitude and most sensitive timing parameter. For the $\pm4\hbar k$ target state ($n=2$), the equal-amplitude square-pulse protocol is significantly more sensitive than the other three protocols to the experimental parameters. For this target state, the triangle pulse protocol has both the highest fidelity  and the lowest sensitivity to experimental parameters.

For the $\pm8\hbar k$ target state ($n=4$), a tuning with a 0.982 fidelity is presented in Tab.\ \ref{table:ue-sensitivity-square}. However, this tuning is exceptionally sensitive to the variation of parameters. Among the other tunings, the sinc-squared and triangle pulses have better fidelity and less sensitivity to the pulse intensity, but are modestly more sensitive to the pulse timing.

\begin{table}
\caption{\label{table:sensitivity-square}
Sensitivity to pulse parameters for equal amplitude square light pulses.}
\begin{tabular}{>{\centering}p{0.55in}
                >{\centering}p{0.35in}
								>{\centering}p{0.05in}
								>{\centering}p{0.35in}
								>{\centering}p{0.4in}
								>{\centering}p{0.4in}
								>{\centering}p{0.4in}
								>{\centering\arraybackslash}p{0.4in}}
\hline
\multicolumn{2}{c}{Target State} & ~~ &%
\multicolumn{5}{c}{Sensitivity Values} \\[0.06in]
$\displaystyle \pm2n\hbar k_0$  &%
$\displaystyle |C^+_{2n}|^2$ &%
&%
$\displaystyle {\Omega_1}$ &%
$\displaystyle {\Omega_2}$ &%
$\displaystyle {\tau_1}$ &%
$\displaystyle {\tau_2}$ &%
$\displaystyle {\tau_3}$ \\[0.08in] 
\cline{1-2}\cline{4-8} \\[-0.1in]
$\pm2\hbar k_0$    & .999   & & 1.13      & 1.08    & 2.20     & 2.94    & 7.54 \\
$\pm4\hbar k_0$    & .991   & & 69.8      & 26.6    & 146.     & 175.    & 63.9 \\
$\pm6\hbar k_0$    & .968   & & 21.3      & 12.4    & 36.2     & 43.8    & 279. \\
$\pm8\hbar k_0$    & .923   & & 319.      & 12.4    & 449      & 60.9    & 195. \\   
\hline
\end{tabular}
\end{table}

\begin{table}
\caption{\label{table:ue-sensitivity-square}
Sensitivity to pulse parameters for unequal amplitude square light pulses.}
\begin{tabular}{>{\centering}p{0.55in}
                >{\centering}p{0.35in}
								>{\centering}p{0.05in}
								>{\centering}p{0.35in}
								>{\centering}p{0.4in}
								>{\centering}p{0.4in}
								>{\centering}p{0.4in}
								>{\centering\arraybackslash}p{0.4in}}
\hline
\multicolumn{2}{c}{Target State} & ~~ &%
\multicolumn{5}{c}{Sensitivity Values} \\[0.06in]
$\displaystyle \pm2n\hbar k_0$  &%
$\displaystyle |C^+_{2n}|^2$ &%
&%
$\displaystyle {\Omega_1}$ &%
$\displaystyle {\Omega_2}$ &%
$\displaystyle {\tau_1}$ &%
$\displaystyle {\tau_2}$ &%
$\displaystyle {\tau_3}$ \\[0.08in] 
\cline{1-2}\cline{4-8} \\[-0.1in]
$\pm2\hbar k_0$   & .999  & & 1.09       & 1.10   & 3.29    & 4.35  & 4.44 \\
$\pm4\hbar k_0$   & .996  & & 21.2       & 5.62   & 39.1    & 17.0  & 28.1 \\
$\pm6\hbar k_0$   & .975  & & 20.9       & 12.2   & 36.3    & 42.3  & 267. \\
$\pm8\hbar k_0$   & .903  & & 237.       & 57.7   & 444.    & 377.  & 172. \\ 
$\pm8\hbar k_0$   & .943  & & 330.       & 11.7   & 616.    & 213.  & 57.5 \\
$\pm8\hbar k_0$   & .982  & & 3700       & 755.   & 5790    & 3180  & 3520 \\
\hline
\end{tabular}
\end{table}

\begin{table}
\caption{\label{table:sensitivity-sinc-square}
Sensitivity to pulse parameters for sinc-squared light pulses.}
\begin{tabular}{>{\centering}p{0.55in}
                >{\centering}p{0.35in}
								>{\centering}p{0.05in}
								>{\centering}p{0.35in}
								>{\centering}p{0.4in}
								>{\centering}p{0.4in}
								>{\centering}p{0.4in}
								>{\centering\arraybackslash}p{0.4in}}
\hline
\multicolumn{2}{c}{Target State} & ~~ &%
\multicolumn{5}{c}{Sensitivity Values} \\[0.06in]
$\displaystyle \pm2n\hbar k_0$  &%
$\displaystyle |C^+_{2n}|^2$ &%
&%
$\displaystyle {\Omega_1}$ &%
$\displaystyle {\Omega_2}$ &%
$\displaystyle {\tau_1}$ &%
$\displaystyle {\tau_2}$ &%
$\displaystyle {\tau_3}$ \\[0.08in] 
\cline{1-2}\cline{4-8} \\[-0.1in] 
$\pm2\hbar k_0$   & .999  &  & 0.925      & 1.57   & 2.19     & 7.77  & 2.82\\
$\pm4\hbar k_0$   & .958  &  & 24.4       & 6.41   & 71.7     & 12.0  & 1.07\\
$\pm6\hbar k_0$   & .866  &  & 72.8       & 11.9   & 183.     & 9.09  & 15.2\\
$\pm8\hbar k_0$   & .958  &  & 39.5       & 17.0   & 124.     & 71.0  & 498.\\   
\hline
\end{tabular}
\end{table}

\begin{table}
\caption{\label{table:sensitivity-triangle}
Sensitivity to pulse parameters for triangle light pulses.}
\begin{tabular}{>{\centering}p{0.45in}
                >{\centering}p{0.35in}
								>{\centering}p{0.05in}
								>{\centering}p{0.35in}
								>{\centering}p{0.4in}
								>{\centering}p{0.4in}
								>{\centering}p{0.4in}
								>{\centering\arraybackslash}p{0.4in}}
\hline
\multicolumn{2}{c}{Target State} & ~~ &%
\multicolumn{5}{c}{Sensitivity Values} \\[0.06in]
$\displaystyle \pm2n\hbar k_0$  &%
$\displaystyle |C^+_{2n}|^2$ &%
&%
$\displaystyle {\Omega_1}$ &%
$\displaystyle {\Omega_2}$ &%
$\displaystyle {\tau_1}$ &%
$\displaystyle {\tau_2}$ &%
$\displaystyle {\tau_3}$ \\[0.08in] 
\cline{1-2}\cline{4-8} \\[-0.1in] 
$\pm2\hbar k_0$   & .999   & & 1.13       & 1.07   & 3.46      & 4.31    & 2.67\\
$\pm4\hbar k_0$   & .993   & & 9.43       & 13.8   & 19.9      & 25.3    & 38.3\\
$\pm6\hbar k_0$   & .971   & & 22.7       & 12.8   & 61.9      & 38.2    & 160.\\
$\pm8\hbar k_0$   & .963   & & 39.7       & 17.3   & 105.      & 56.2    & 610.\\   
\hline
\end{tabular}
\end{table}

\section{Analysis and Discussion}\label{sec:discussion}

\subsection{Convergence Tests}
The results of the convergence sensitivity tests for the number of slices and the number of retained coefficients demonstrate the validity of our numerical method. Convergence was obtained with modest values of $N_\mathrm{s}$ and $N$. The calculations were carried out with $N_\mathrm{s}=50$ slices per pulse and $N=24$ retained coefficients without taxing our computational resources. The validity of the numerical method is also demonstrated by benchmarking against previously published calculations for the equal-amplitude square-pulse protocol.\cite{wu2005splitting} 

\subsection{Equal and Varied Amplitude Square Pulses}
Equal-amplitude two-pulse protocols utilizing square pulses allow highly selective excitation of beam-splitter states, which is not possible with earlier single-pulse protocols.\cite{wu2005splitting} While square two-pulse protocols with equal amplitude are effective, we investigate whether setting the intensity of the pulses independently results in better fidelity  or reduced sensitivity to experimental parameters, especially in the case of higher-order target states. While the resulting change in fidelity  is not significant, the protocol with unequal pulse intensities is significantly less sensitive to changes in parameters, especially with regard to variations in pulse intensity. This improvement suggests that additional gains could be achieved by varying other pulse features such as shape. 

\subsection{Sinc-Squared and Triangle-Shaped Pulses}
Fidelity  and sensitivity were also assessed for beam-splitter protocols utilizing sinc-squared and triangle-shaped pulses. 
The shaped pulses showed similar results for the first two momentum states, with the triangle-shaped pulses showing higher fidelity for any given state. 
The sinc-squared and triangle pulses show a significant increase in fidelity in the $\pm8\hbar k$ state when compared to the square pulses of comparable parameter sensitivity, demonstrating that pulse envelope shape does affect the fidelity and sensitivity for high-order momentum states. Parameter sensitivity analysis of the triangle and sinc-squared pulses also demonstrates the potential for significantly reduced sensitivity to optical-pulse amplitude for the $\pm4\hbar k$ and $\pm8\hbar k$ target states relative to square-shaped pulses. Low sensitivity to pulse amplitude is particularly important due to the technical challenges of controlling absolute pulse intensities experimentally.

\subsection{High Fidelity and Low Sensitivity}
\begin{figure*}
\includegraphics[width=6.69in]{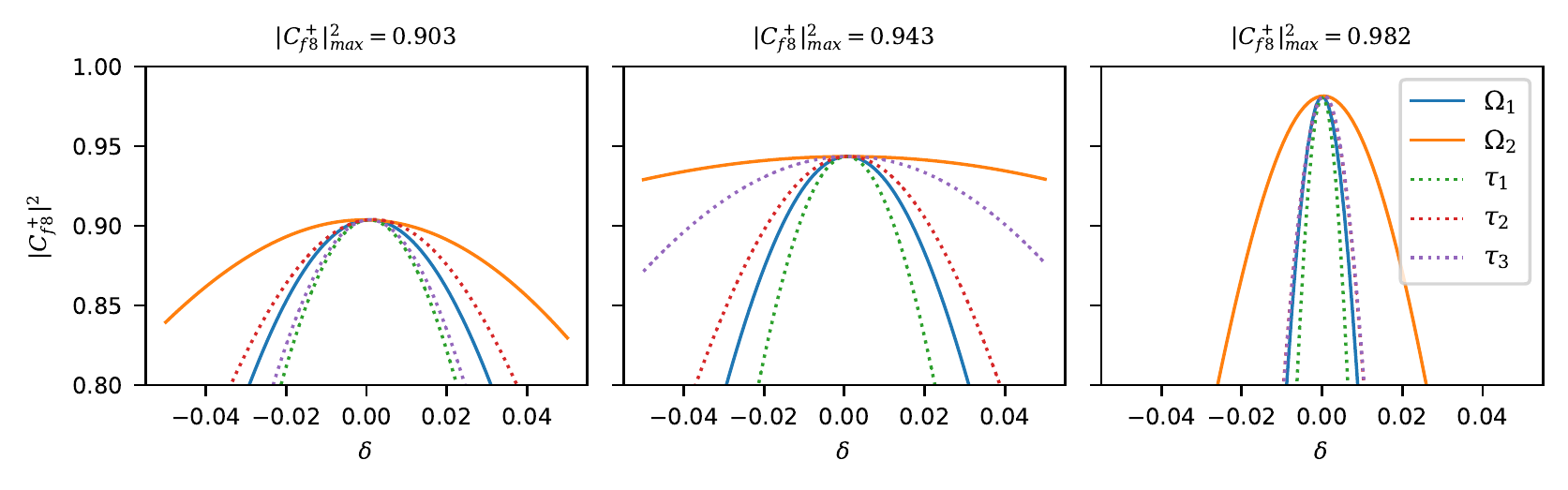}
 \caption{\label{Sensitivity_Fidelity_8hbark}
Sensitivity of the fidelity for the $\pm8\hbar k$ states to variations in pulse parameters for different tunings of unequal square pulses. The three tunings demonstrate there are multiple local maximums of the fidelity in the parameter space. The high sensitivity of the tuning with 98\% fidelity illustrates the need to consider trade-offs between fidelity and sensitivity to pulse parameter fluctuations.}
\end{figure*}

The beam-splitter protocols used in this paper consider not only high fidelity, but also the sensitivity of the optimized parameters to the final-state fidelity. Target-state fidelity is a complicated function of the pulse parameters, and many locally optimized tunings exist in parameter space. The optimized parameter tunings that we report are representative and do not necessarily correspond to global fidelity maximums.
For example, the first optimal solution listed in Tab.~\ref{table:unequal-amp-square-pulses} for the $\pm8\hbar k$ state using unequal amplitude square pulses, $|C^+_{8}|^2 = 0.903$, is not the highest fidelity found for the pulse shape. Higher-fidelity tunings exist with $|C^+_{8}|^2 = 0.943$ and $|C^+_{8}|^2 = 0.982$. However, these tunings have higher optimal-parameter sensitivity values as shown in Tab.~\ref{table:ue-sensitivity-square} and Fig.~\ref{Sensitivity_Fidelity_8hbark}.

In extending the analysis of matter-wave beam splitters to include both high fidelity and low sensitivity of parameters to final-state fidelity, we present acceptable fidelity values with the lower sensitivity values, some of which have sensitivities an order of magnitude lower than previously presented equal amplitude square pulse sequences. Acceptable fidelity at lower sensitivity may be necessary for robust experimental realization of these protocols.

As a practical matter, precision timing is much easier to achieve experimentally than is precision optical-pulse intensity. Restricting consideration of parameter sensitivity to intensity variations, we find the following general trends for the choice of pulse shape: for the $\pm2\hbar k_0$ state, the pulse shapes considered are roughly equivalent; for the $\pm4\hbar k_0$ state, the triangle pulses are the best choice; for the $\pm6\hbar k_0$ state, the sinc-squared pulse shape is significantly worse than square or triangular pulses; and for the $\pm8\hbar k_0$ state, the sinc-squared and triangle pulses are equivalent and both are superior to the square pulses.  

\subsection{Non-Zero Initial Momentum.}We have focused our efforts on optimizing optical pulses for the case $\hbar k = 0$ of atoms with zero initial momentum. Whether due to variations in a statistical ensemble or the finite extent of the atom wave packet, in atom interferometry experiments, the atoms are initially in a distribution of momentum states centered about zero. Accordingly, whether optical pulses optimized for zero initial momentum maintain good target state fidelity over an achievable range of initial momentum is an important practical question.

\begin{figure}
\includegraphics{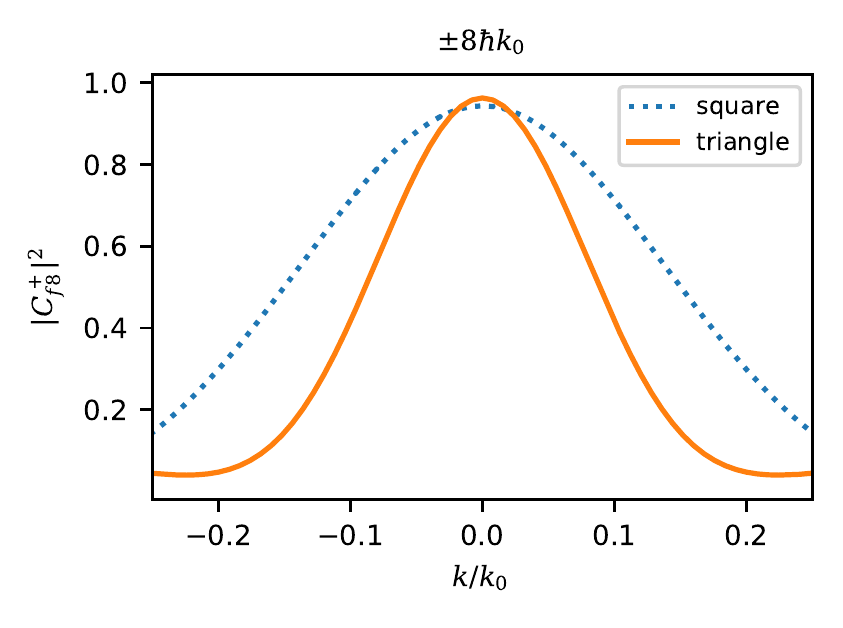}
\caption{\label{k_sensitivity}Sensitivity of fidelity to initial momentum for the $\pm8\hbar k_0$ state for square and triangle pulse shapes. Initial momentum $\hbar k$ is plotted as a fraction of the recoil momentum $\hbar k_0$. Here, the fidelity is more sensitive to $k$ for the triangle pulses than it is for the square pulses; however, for $k/k_0<0.02$, the fidelity for the triangle pulses exceeds the fidelity for the square pulses. The worst case is shown. For $\pm2\hbar k_0$, $\pm4\hbar k_0$, and $\pm6\hbar k_0$ the sensitivity of the fidelity to $k$ for the triangle and sinc-squared pulses is less than or equal to the sensitivity for the corresponding square pulses.}
\end{figure}

To answer this question, we solve Eq.\  (\ref{eq:one_atom_SE}) for various values of $k$ using methods similar to those described in Sec. \ref{sec:method} and keeping the optimal pulse parameters calculated for $k=0$. Representative results are plotted in Fig.\ \ref{k_sensitivity}. We find that for the $\pm2\hbar k_0$ and $\pm6\hbar k_0$ states, the sensitivity of the fidelity to $k$ is roughly the same for all of the pulse shapes we considered. For the $\pm4\hbar k_0$ state, the fidelity for the triangle and sinc-squared pulses is less sensitive to $k$ than the fidelity for the square pulses. For the $\pm8\hbar k_0$ state, the fidelity is more sensitive to $k$ for the triangle and sinc-squared pulses. Even so, the fidelity for the triangle and sinc-squared pulses is higher than the fidelity for the experimentally achievable range of $|k|<0.02k_0$.

\section{Conclusion and Future Work}\label{sec:conclusion}
In this paper, we show how varying the shape of optical pulses used to split BECs in atom interferometry improves optimal beam-splitter fidelity and reduces sensitivity to optimized parameters. In some cases, we have reduced parameter sensitivity by an order of magnitude while preserving fidelity.
The results of this work show that using shaped optical pulses has beneficial effects in the excitation of high fidelity beam-splitter states in BECs, at least under conditions where the one-dimensional, non-interacting, and transitionally invariant approximations are valid.

Our analysis relies on brute force to identify optimal pulse sequences. Appeal to an approximate Bloch-sphere description provides a conceptual basis for pulse sequences that excite the $n=1$ beam-splitter state, but a comparable understanding of the excitation of higher-order states is lacking, and our explorations have not yielded additional insight.\cite{wu2005splitting} In the absence of a conceptual model, there is a concern that the improvements in sensitivity to experimental parameters, particularly for excitation with triangular optical pulses, might depend on the sharp corners of the pulse shape, which are idealizations that are difficult to replicate experimentally. However, such dependence is unlikely for several reasons. First, the solutions of the Schr{\"o}dinger equation are not sensitive to finite temporal discontinuities in the potential or its first derivatives, and we do not observe any anomalous behavior of our solutions in the vicinity of the sharp corners. Second, the triangular pulses, while more technically demanding to generate, are actually less-severe approximations to experimentally obtainable pulses than are the previously studied and experimentally demonstrated square pulses. Finally, the rapid convergence of our calculations with the number of temporal slices used demonstrates a lack of sensitivity of the results to details of the pulses on very short time scales.

In addition to seeking a conceptual model for excitation of high momentum beam-splitter states, future work will explore the possibility of systematic design of pulse sequences with an emphasis on the significance of the post-first-pulse state in setting the stage for selective excitation of high-order beam-splitter states.
Additionally, if our general predictions are  validated experimentally, we will extend our analysis to 3d models that incorporate atom--atom interactions
and translational variation in atom density. 
Finally, we would like to explore the application of our optimization procedures to a broader class of pulse envelope shapes, or more generally to envelopes attained by optimizing the pulse amplitude in each time slice.


\section*{Conflict of interest}
The authors have no conflicts to disclose.

\section*{Data availability}
Data sharing is not applicable to this article as no new data were created or analyzed in this study.

\section*{Acknowledgements}
This work was funded in part by the United States Military Academy Department of Electrical Engineering and Computer Science and the Department of Physics and Nuclear Engineering. We thank Kirk Ingold (United States Military Academy) and Corey Gerving (United States Military Academy) for helpful discussion that contributed to the success of this project. MCC would like to thank Service Academies Research Associates (SARRA) program at LANL, funded by the National Nuclear Security Program’s Military Academy Collaboration (MAC). MGB acknowledges support from the DARPA MTO A-PhI Program and the use of code developed under award 20180045DR from the Laboratory Directed Research and Development program of Los Alamos National Laboratory. 

The conclusions of this work are those of the authors and do not reflect official positions of the Department of the Army or the Department of Defense, or the United States government. As employees of the U.S. Government, the authors’ copyright interest are subject to limitations under U.S. copyright law. This manuscript has been cleared for public release.

\bibliography{CassidyBoshierHarrell2021.bib}

\begin{thebibliography}{24}%
\makeatletter
\providecommand \@ifxundefined [1]{%
 \@ifx{#1\undefined}
}%
\providecommand \@ifnum [1]{%
 \ifnum #1\expandafter \@firstoftwo
 \else \expandafter \@secondoftwo
 \fi
}%
\providecommand \@ifx [1]{%
 \ifx #1\expandafter \@firstoftwo
 \else \expandafter \@secondoftwo
 \fi
}%
\providecommand \natexlab [1]{#1}%
\providecommand \enquote  [1]{``#1''}%
\providecommand \bibnamefont  [1]{#1}%
\providecommand \bibfnamefont [1]{#1}%
\providecommand \citenamefont [1]{#1}%
\providecommand \href@noop [0]{\@secondoftwo}%
\providecommand \href [0]{\begingroup \@sanitize@url \@href}%
\providecommand \@href[1]{\@@startlink{#1}\@@href}%
\providecommand \@@href[1]{\endgroup#1\@@endlink}%
\providecommand \@sanitize@url [0]{\catcode `\\12\catcode `\$12\catcode
  `\&12\catcode `\#12\catcode `\^12\catcode `\_12\catcode `\%12\relax}%
\providecommand \@@startlink[1]{}%
\providecommand \@@endlink[0]{}%
\providecommand \url  [0]{\begingroup\@sanitize@url \@url }%
\providecommand \@url [1]{\endgroup\@href {#1}{\urlprefix }}%
\providecommand \urlprefix  [0]{URL }%
\providecommand \Eprint [0]{\href }%
\providecommand \doibase [0]{http://dx.doi.org/}%
\providecommand \selectlanguage [0]{\@gobble}%
\providecommand \bibinfo  [0]{\@secondoftwo}%
\providecommand \bibfield  [0]{\@secondoftwo}%
\providecommand \translation [1]{[#1]}%
\providecommand \BibitemOpen [0]{}%
\providecommand \bibitemStop [0]{}%
\providecommand \bibitemNoStop [0]{.\EOS\space}%
\providecommand \EOS [0]{\spacefactor3000\relax}%
\providecommand \BibitemShut  [1]{\csname bibitem#1\endcsname}%
\let\auto@bib@innerbib\@empty
\bibitem [{\citenamefont {Cronin}, \citenamefont {Schmiedmayer},\ and\
  \citenamefont {Pritchard}(2009)}]{RevModPhys.81.1051}%
  \BibitemOpen
  \bibfield  {author} {\bibinfo {author} {\bibfnamefont {A.~D.}\ \bibnamefont
  {Cronin}}, \bibinfo {author} {\bibfnamefont {J.}~\bibnamefont
  {Schmiedmayer}}, \ and\ \bibinfo {author} {\bibfnamefont {D.~E.}\
  \bibnamefont {Pritchard}},\ }\href {\doibase 10.1103/RevModPhys.81.1051}
  {\bibfield  {journal} {\bibinfo  {journal} {Rev. Mod. Phys.}\ }\textbf
  {\bibinfo {volume} {81}},\ \bibinfo {pages} {1051} (\bibinfo {year}
  {2009})}\BibitemShut {NoStop}%
\bibitem [{\citenamefont {Wang}\ \emph {et~al.}(2005)\citenamefont {Wang},
  \citenamefont {Anderson}, \citenamefont {Bright}, \citenamefont {Cornell},
  \citenamefont {Diot}, \citenamefont {Kishimoto}, \citenamefont {Prentiss},
  \citenamefont {Saravanan}, \citenamefont {Segal},\ and\ \citenamefont
  {Wu}}]{wang2005atom}%
  \BibitemOpen
  \bibfield  {author} {\bibinfo {author} {\bibfnamefont {Y.-J.}\ \bibnamefont
  {Wang}}, \bibinfo {author} {\bibfnamefont {D.~Z.}\ \bibnamefont {Anderson}},
  \bibinfo {author} {\bibfnamefont {V.~M.}\ \bibnamefont {Bright}}, \bibinfo
  {author} {\bibfnamefont {E.~A.}\ \bibnamefont {Cornell}}, \bibinfo {author}
  {\bibfnamefont {Q.}~\bibnamefont {Diot}}, \bibinfo {author} {\bibfnamefont
  {T.}~\bibnamefont {Kishimoto}}, \bibinfo {author} {\bibfnamefont
  {M.}~\bibnamefont {Prentiss}}, \bibinfo {author} {\bibfnamefont {R.~A.}\
  \bibnamefont {Saravanan}}, \bibinfo {author} {\bibfnamefont {S.~R.}\
  \bibnamefont {Segal}}, \ and\ \bibinfo {author} {\bibfnamefont
  {S.}~\bibnamefont {Wu}},\ }\href {\doibase 10.1103/PhysRevLett.94.090405}
  {\bibfield  {journal} {\bibinfo  {journal} {Phys. Rev. Lett.}\ }\textbf
  {\bibinfo {volume} {94}},\ \bibinfo {pages} {090405} (\bibinfo {year}
  {2005})}\BibitemShut {NoStop}%
\bibitem [{\citenamefont {Garcia}\ \emph {et~al.}(2006)\citenamefont {Garcia},
  \citenamefont {Deissler}, \citenamefont {Hughes}, \citenamefont {Reeves},\
  and\ \citenamefont {Sackett}}]{garcia2006bose}%
  \BibitemOpen
  \bibfield  {author} {\bibinfo {author} {\bibfnamefont {O.}~\bibnamefont
  {Garcia}}, \bibinfo {author} {\bibfnamefont {B.}~\bibnamefont {Deissler}},
  \bibinfo {author} {\bibfnamefont {K.~J.}\ \bibnamefont {Hughes}}, \bibinfo
  {author} {\bibfnamefont {J.~M.}\ \bibnamefont {Reeves}}, \ and\ \bibinfo
  {author} {\bibfnamefont {C.~A.}\ \bibnamefont {Sackett}},\ }\href {\doibase
  10.1103/PhysRevA.74.031601} {\bibfield  {journal} {\bibinfo  {journal} {Phys.
  Rev. A}\ }\textbf {\bibinfo {volume} {74}},\ \bibinfo {pages} {031601}
  (\bibinfo {year} {2006})}\BibitemShut {NoStop}%
\bibitem [{\citenamefont {Burke}\ \emph {et~al.}(2008)\citenamefont {Burke},
  \citenamefont {Deissler}, \citenamefont {Hughes},\ and\ \citenamefont
  {Sackett}}]{burke2008confinement}%
  \BibitemOpen
  \bibfield  {author} {\bibinfo {author} {\bibfnamefont {J.~H.~T.}\
  \bibnamefont {Burke}}, \bibinfo {author} {\bibfnamefont {B.}~\bibnamefont
  {Deissler}}, \bibinfo {author} {\bibfnamefont {K.~J.}\ \bibnamefont
  {Hughes}}, \ and\ \bibinfo {author} {\bibfnamefont {C.~A.}\ \bibnamefont
  {Sackett}},\ }\href {\doibase 10.1103/PhysRevA.78.023619} {\bibfield
  {journal} {\bibinfo  {journal} {Phys. Rev. A}\ }\textbf {\bibinfo {volume}
  {78}},\ \bibinfo {pages} {023619} (\bibinfo {year} {2008})}\BibitemShut
  {NoStop}%
\bibitem [{\citenamefont {Bongs}\ \emph {et~al.}(2019)\citenamefont {Bongs},
  \citenamefont {Holynski}, \citenamefont {Vovrosh}, \citenamefont {Bouyer},
  \citenamefont {Condon}, \citenamefont {Rasel}, \citenamefont {Schubert},
  \citenamefont {Schleich},\ and\ \citenamefont {Roura}}]{bongs2019taking}%
  \BibitemOpen
  \bibfield  {author} {\bibinfo {author} {\bibfnamefont {K.}~\bibnamefont
  {Bongs}}, \bibinfo {author} {\bibfnamefont {M.}~\bibnamefont {Holynski}},
  \bibinfo {author} {\bibfnamefont {J.}~\bibnamefont {Vovrosh}}, \bibinfo
  {author} {\bibfnamefont {P.}~\bibnamefont {Bouyer}}, \bibinfo {author}
  {\bibfnamefont {G.}~\bibnamefont {Condon}}, \bibinfo {author} {\bibfnamefont
  {E.}~\bibnamefont {Rasel}}, \bibinfo {author} {\bibfnamefont
  {C.}~\bibnamefont {Schubert}}, \bibinfo {author} {\bibfnamefont {W.~P.}\
  \bibnamefont {Schleich}}, \ and\ \bibinfo {author} {\bibfnamefont
  {A.}~\bibnamefont {Roura}},\ }\href {\doibase
  https://doi.org/10.1038/s42254-019-0117-4} {\bibfield  {journal} {\bibinfo
  {journal} {Nat. Rev. Phys.}\ }\textbf {\bibinfo {volume} {1}},\ \bibinfo
  {pages} {731} (\bibinfo {year} {2019})}\BibitemShut {NoStop}%
\bibitem [{\citenamefont {Wu}\ \emph {et~al.}(2005)\citenamefont {Wu},
  \citenamefont {Wang}, \citenamefont {Diot},\ and\ \citenamefont
  {Prentiss}}]{wu2005splitting}%
  \BibitemOpen
  \bibfield  {author} {\bibinfo {author} {\bibfnamefont {S.}~\bibnamefont
  {Wu}}, \bibinfo {author} {\bibfnamefont {Y.-J.}\ \bibnamefont {Wang}},
  \bibinfo {author} {\bibfnamefont {Q.}~\bibnamefont {Diot}}, \ and\ \bibinfo
  {author} {\bibfnamefont {M.}~\bibnamefont {Prentiss}},\ }\href {\doibase
  10.1103/PhysRevA.71.043602} {\bibfield  {journal} {\bibinfo  {journal} {Phys.
  Rev. A}\ }\textbf {\bibinfo {volume} {71}},\ \bibinfo {pages} {043602}
  (\bibinfo {year} {2005})}\BibitemShut {NoStop}%
\bibitem [{\citenamefont {Stickney}\ \emph {et~al.}(2008)\citenamefont
  {Stickney}, \citenamefont {Kafle}, \citenamefont {Anderson},\ and\
  \citenamefont {Zozulya}}]{stickney2008theoretical}%
  \BibitemOpen
  \bibfield  {author} {\bibinfo {author} {\bibfnamefont {J.~A.}\ \bibnamefont
  {Stickney}}, \bibinfo {author} {\bibfnamefont {R.~P.}\ \bibnamefont {Kafle}},
  \bibinfo {author} {\bibfnamefont {D.~Z.}\ \bibnamefont {Anderson}}, \ and\
  \bibinfo {author} {\bibfnamefont {A.~A.}\ \bibnamefont {Zozulya}},\ }\href
  {\doibase 10.1103/PhysRevA.77.043604} {\bibfield  {journal} {\bibinfo
  {journal} {Phys. Rev. A}\ }\textbf {\bibinfo {volume} {77}},\ \bibinfo
  {pages} {043604} (\bibinfo {year} {2008})}\BibitemShut {NoStop}%
\bibitem [{\citenamefont {Edwards}\ \emph {et~al.}(2010)\citenamefont
  {Edwards}, \citenamefont {Benton}, \citenamefont {Heward},\ and\
  \citenamefont {Clark}}]{edwards2010momentum}%
  \BibitemOpen
  \bibfield  {author} {\bibinfo {author} {\bibfnamefont {M.}~\bibnamefont
  {Edwards}}, \bibinfo {author} {\bibfnamefont {B.}~\bibnamefont {Benton}},
  \bibinfo {author} {\bibfnamefont {J.}~\bibnamefont {Heward}}, \ and\ \bibinfo
  {author} {\bibfnamefont {C.~W.}\ \bibnamefont {Clark}},\ }\href {\doibase
  10.1103/PhysRevA.82.063613} {\bibfield  {journal} {\bibinfo  {journal} {Phys.
  Rev. A}\ }\textbf {\bibinfo {volume} {82}},\ \bibinfo {pages} {063613}
  (\bibinfo {year} {2010})}\BibitemShut {NoStop}%
\bibitem [{\citenamefont {Crookston}, \citenamefont {Baker},\ and\
  \citenamefont {Robinson}(2005)}]{crookston2005microchip}%
  \BibitemOpen
  \bibfield  {author} {\bibinfo {author} {\bibfnamefont {M.~B.}\ \bibnamefont
  {Crookston}}, \bibinfo {author} {\bibfnamefont {P.~M.}\ \bibnamefont
  {Baker}}, \ and\ \bibinfo {author} {\bibfnamefont {M.~P.}\ \bibnamefont
  {Robinson}},\ }\href {\doibase 10.1088/0953-4075/38/18/001} {\bibfield
  {journal} {\bibinfo  {journal} {J. Phys. B: At. Mol. Opt. Phys.}\ }\textbf
  {\bibinfo {volume} {38}},\ \bibinfo {pages} {3289} (\bibinfo {year}
  {2005})}\BibitemShut {NoStop}%
\bibitem [{\citenamefont {Bord\'e}(1995)}]{borde1995amplification}%
  \BibitemOpen
  \bibfield  {author} {\bibinfo {author} {\bibfnamefont {C.}~\bibnamefont
  {Bord\'e}},\ }\href {\doibase https://doi.org/10.1016/0375-9601(95)00466-G}
  {\bibfield  {journal} {\bibinfo  {journal} {Phys. Lett. A}\ }\textbf
  {\bibinfo {volume} {204}},\ \bibinfo {pages} {217} (\bibinfo {year}
  {1995})}\BibitemShut {NoStop}%
\bibitem [{\citenamefont {Dalfovo}\ \emph {et~al.}(1999)\citenamefont
  {Dalfovo}, \citenamefont {Giorgini}, \citenamefont {Pitaevskii},\ and\
  \citenamefont {Stringari}}]{dalfovo1999theory}%
  \BibitemOpen
  \bibfield  {author} {\bibinfo {author} {\bibfnamefont {F.}~\bibnamefont
  {Dalfovo}}, \bibinfo {author} {\bibfnamefont {S.}~\bibnamefont {Giorgini}},
  \bibinfo {author} {\bibfnamefont {L.~P.}\ \bibnamefont {Pitaevskii}}, \ and\
  \bibinfo {author} {\bibfnamefont {S.}~\bibnamefont {Stringari}},\ }\href
  {\doibase 10.1103/RevModPhys.71.463} {\bibfield  {journal} {\bibinfo
  {journal} {Rev. Mod. Phys.}\ }\textbf {\bibinfo {volume} {71}},\ \bibinfo
  {pages} {463} (\bibinfo {year} {1999})}\BibitemShut {NoStop}%
\bibitem [{\citenamefont {Jamison}, \citenamefont {Kutz},\ and\ \citenamefont
  {Gupta}(2011)}]{jamison2011atomic}%
  \BibitemOpen
  \bibfield  {author} {\bibinfo {author} {\bibfnamefont {A.~O.}\ \bibnamefont
  {Jamison}}, \bibinfo {author} {\bibfnamefont {J.~N.}\ \bibnamefont {Kutz}}, \
  and\ \bibinfo {author} {\bibfnamefont {S.}~\bibnamefont {Gupta}},\ }\href
  {\doibase 10.1103/PhysRevA.84.043643} {\bibfield  {journal} {\bibinfo
  {journal} {Phys. Rev. A}\ }\textbf {\bibinfo {volume} {84}},\ \bibinfo
  {pages} {043643} (\bibinfo {year} {2011})}\BibitemShut {NoStop}%
\bibitem [{\citenamefont {Gould}, \citenamefont {Ruff},\ and\ \citenamefont
  {Pritchard}(1986)}]{gould1986diffraction}%
  \BibitemOpen
  \bibfield  {author} {\bibinfo {author} {\bibfnamefont {P.~L.}\ \bibnamefont
  {Gould}}, \bibinfo {author} {\bibfnamefont {G.~A.}\ \bibnamefont {Ruff}}, \
  and\ \bibinfo {author} {\bibfnamefont {D.~E.}\ \bibnamefont {Pritchard}},\
  }\href {\doibase 10.1103/PhysRevLett.56.827} {\bibfield  {journal} {\bibinfo
  {journal} {Phys. Rev. Lett.}\ }\textbf {\bibinfo {volume} {56}},\ \bibinfo
  {pages} {827} (\bibinfo {year} {1986})}\BibitemShut {NoStop}%
\bibitem [{\citenamefont {Xiong}\ \emph {et~al.}(2011)\citenamefont {Xiong},
  \citenamefont {Yue}, \citenamefont {Wang}, \citenamefont {Zhou},\ and\
  \citenamefont {Chen}}]{xiong2011manipulating}%
  \BibitemOpen
  \bibfield  {author} {\bibinfo {author} {\bibfnamefont {W.}~\bibnamefont
  {Xiong}}, \bibinfo {author} {\bibfnamefont {X.}~\bibnamefont {Yue}}, \bibinfo
  {author} {\bibfnamefont {Z.}~\bibnamefont {Wang}}, \bibinfo {author}
  {\bibfnamefont {X.}~\bibnamefont {Zhou}}, \ and\ \bibinfo {author}
  {\bibfnamefont {X.}~\bibnamefont {Chen}},\ }\href {\doibase
  10.1103/PhysRevA.84.043616} {\bibfield  {journal} {\bibinfo  {journal} {Phys.
  Rev. A}\ }\textbf {\bibinfo {volume} {84}},\ \bibinfo {pages} {043616}
  (\bibinfo {year} {2011})}\BibitemShut {NoStop}%
\bibitem [{\citenamefont {Wu}, \citenamefont {Su},\ and\ \citenamefont
  {Prentiss}(2007)}]{wu2007demonstration}%
  \BibitemOpen
  \bibfield  {author} {\bibinfo {author} {\bibfnamefont {S.}~\bibnamefont
  {Wu}}, \bibinfo {author} {\bibfnamefont {E.}~\bibnamefont {Su}}, \ and\
  \bibinfo {author} {\bibfnamefont {M.}~\bibnamefont {Prentiss}},\ }\href
  {\doibase 10.1103/PhysRevLett.99.173201} {\bibfield  {journal} {\bibinfo
  {journal} {Phys. Rev. Lett.}\ }\textbf {\bibinfo {volume} {99}},\ \bibinfo
  {pages} {173201} (\bibinfo {year} {2007})}\BibitemShut {NoStop}%
\bibitem [{\citenamefont {Hughes}\ \emph {et~al.}(2007)\citenamefont {Hughes},
  \citenamefont {Deissler}, \citenamefont {Burke},\ and\ \citenamefont
  {Sackett}}]{hughes2007high}%
  \BibitemOpen
  \bibfield  {author} {\bibinfo {author} {\bibfnamefont {K.~J.}\ \bibnamefont
  {Hughes}}, \bibinfo {author} {\bibfnamefont {B.}~\bibnamefont {Deissler}},
  \bibinfo {author} {\bibfnamefont {J.~H.~T.}\ \bibnamefont {Burke}}, \ and\
  \bibinfo {author} {\bibfnamefont {C.~A.}\ \bibnamefont {Sackett}},\ }\href
  {\doibase 10.1103/PhysRevA.76.035601} {\bibfield  {journal} {\bibinfo
  {journal} {Phys. Rev. A}\ }\textbf {\bibinfo {volume} {76}},\ \bibinfo
  {pages} {035601} (\bibinfo {year} {2007})}\BibitemShut {NoStop}%
\bibitem [{\citenamefont {M{\"u}ller}, \citenamefont {Chiow},\ and\
  \citenamefont {Chu}(2008)}]{muller2008atom}%
  \BibitemOpen
  \bibfield  {author} {\bibinfo {author} {\bibfnamefont {H.}~\bibnamefont
  {M{\"u}ller}}, \bibinfo {author} {\bibfnamefont {S.-w.}\ \bibnamefont
  {Chiow}}, \ and\ \bibinfo {author} {\bibfnamefont {S.}~\bibnamefont {Chu}},\
  }\href {\doibase 10.1103/PhysRevA.77.023609} {\bibfield  {journal} {\bibinfo
  {journal} {Phys. Rev. A}\ }\textbf {\bibinfo {volume} {77}},\ \bibinfo
  {pages} {023609} (\bibinfo {year} {2008})}\BibitemShut {NoStop}%
\bibitem [{\citenamefont {Gadway}\ \emph {et~al.}(2009)\citenamefont {Gadway},
  \citenamefont {Pertot}, \citenamefont {Reimann}, \citenamefont {Cohen},\ and\
  \citenamefont {Schneble}}]{gadway2009analysis}%
  \BibitemOpen
  \bibfield  {author} {\bibinfo {author} {\bibfnamefont {B.}~\bibnamefont
  {Gadway}}, \bibinfo {author} {\bibfnamefont {D.}~\bibnamefont {Pertot}},
  \bibinfo {author} {\bibfnamefont {R.}~\bibnamefont {Reimann}}, \bibinfo
  {author} {\bibfnamefont {M.~G.}\ \bibnamefont {Cohen}}, \ and\ \bibinfo
  {author} {\bibfnamefont {D.}~\bibnamefont {Schneble}},\ }\href {\doibase
  10.1364/OE.17.019173} {\bibfield  {journal} {\bibinfo  {journal} {Opt.
  Express}\ }\textbf {\bibinfo {volume} {17}},\ \bibinfo {pages} {19173}
  (\bibinfo {year} {2009})}\BibitemShut {NoStop}%
\bibitem [{\citenamefont {Ammann}\ and\ \citenamefont
  {Christensen}(1997)}]{ammann1997delta}%
  \BibitemOpen
  \bibfield  {author} {\bibinfo {author} {\bibfnamefont {H.}~\bibnamefont
  {Ammann}}\ and\ \bibinfo {author} {\bibfnamefont {N.}~\bibnamefont
  {Christensen}},\ }\href {\doibase 10.1103/PhysRevLett.78.2088} {\bibfield
  {journal} {\bibinfo  {journal} {Phys, Rev. Lett.}\ }\textbf {\bibinfo
  {volume} {78}},\ \bibinfo {pages} {2088} (\bibinfo {year}
  {1997})}\BibitemShut {NoStop}%
\bibitem [{\citenamefont {Meystre}(2001)}]{meystre2001atom}%
  \BibitemOpen
  \bibfield  {author} {\bibinfo {author} {\bibfnamefont {P.}~\bibnamefont
  {Meystre}},\ }\href {https://www.springer.com/us/book/9780387952741} {\emph
  {\bibinfo {title} {Atom Optics}}},\ edited by\ \bibinfo {editor}
  {\bibfnamefont {G.~F.}\ \bibnamefont {Drake}}\ and\ \bibinfo {editor}
  {\bibfnamefont {D.~G.}\ \bibnamefont {Ecker}},\ \bibinfo {series} {Springer
  Series on Atomic, Optical and Plasma Physics}, Vol.~\bibinfo {volume} {33}\
  (\bibinfo  {publisher} {Springer-Verlag, New York},\ \bibinfo {year} {2001})\
  pp.\ \bibinfo {pages} {62, 177}\BibitemShut {NoStop}%
\bibitem [{\citenamefont {Harris}\ \emph {et~al.}(2020)\citenamefont {Harris},
  \citenamefont {Millman}, \citenamefont {van~der Walt}, \citenamefont
  {Gommers}, \citenamefont {Virtanen}, \citenamefont {Cournapeau},
  \citenamefont {Wieser}, \citenamefont {Taylor}, \citenamefont {Berg},
  \citenamefont {Smith} \emph {et~al.}}]{harris2020array}%
  \BibitemOpen
  \bibfield  {author} {\bibinfo {author} {\bibfnamefont {C.~R.}\ \bibnamefont
  {Harris}}, \bibinfo {author} {\bibfnamefont {K.~J.}\ \bibnamefont {Millman}},
  \bibinfo {author} {\bibfnamefont {S.~J.}\ \bibnamefont {van~der Walt}},
  \bibinfo {author} {\bibfnamefont {R.}~\bibnamefont {Gommers}}, \bibinfo
  {author} {\bibfnamefont {P.}~\bibnamefont {Virtanen}}, \bibinfo {author}
  {\bibfnamefont {D.}~\bibnamefont {Cournapeau}}, \bibinfo {author}
  {\bibfnamefont {E.}~\bibnamefont {Wieser}}, \bibinfo {author} {\bibfnamefont
  {J.}~\bibnamefont {Taylor}}, \bibinfo {author} {\bibfnamefont
  {S.}~\bibnamefont {Berg}}, \bibinfo {author} {\bibfnamefont {N.~J.}\
  \bibnamefont {Smith}},  \emph {et~al.},\ }\href {\doibase
  10.1038/s41586-020-2649-2} {\bibfield  {journal} {\bibinfo  {journal}
  {Nature}\ }\textbf {\bibinfo {volume} {585}},\ \bibinfo {pages} {357}
  (\bibinfo {year} {2020})}\BibitemShut {NoStop}%
\bibitem [{\citenamefont {{Virtanen}}\ \emph {et~al.}(2020)\citenamefont
  {{Virtanen}}, \citenamefont {{Gommers}}, \citenamefont {{Oliphant}},
  \citenamefont {{Haberland}}, \citenamefont {{Reddy}}, \citenamefont
  {{Cournapeau}}, \citenamefont {{Burovski}}, \citenamefont {{Peterson}},
  \citenamefont {{Weckesser}}, \citenamefont {{Bright}}, \citenamefont {{van
  der Walt}}, \citenamefont {{Brett}}, \citenamefont {{Wilson}}, \citenamefont
  {{Jarrod Millman}}, \citenamefont {{Mayorov}}, \citenamefont {{Nelson}},
  \citenamefont {{Jones}}, \citenamefont {{Kern}}, \citenamefont {{Larson}},
  \citenamefont {{Carey}}, \citenamefont {{Polat}}, \citenamefont {{Feng}},
  \citenamefont {{Moore}}, \citenamefont {{Vand erPlas}}, \citenamefont
  {{Laxalde}}, \citenamefont {{Perktold}}, \citenamefont {{Cimrman}},
  \citenamefont {{Henriksen}}, \citenamefont {{Quintero}}, \citenamefont
  {{Harris}}, \citenamefont {{Archibald}}, \citenamefont {{Ribeiro}},
  \citenamefont {{Pedregosa}}, \citenamefont {{van Mulbregt}},\ and\
  \citenamefont {{Contributors}}}]{2020SciPy-NMeth}%
  \BibitemOpen
  \bibfield  {author} {\bibinfo {author} {\bibfnamefont {P.}~\bibnamefont
  {{Virtanen}}}, \bibinfo {author} {\bibfnamefont {R.}~\bibnamefont
  {{Gommers}}}, \bibinfo {author} {\bibfnamefont {T.~E.}\ \bibnamefont
  {{Oliphant}}}, \bibinfo {author} {\bibfnamefont {M.}~\bibnamefont
  {{Haberland}}}, \bibinfo {author} {\bibfnamefont {T.}~\bibnamefont
  {{Reddy}}}, \bibinfo {author} {\bibfnamefont {D.}~\bibnamefont
  {{Cournapeau}}}, \bibinfo {author} {\bibfnamefont {E.}~\bibnamefont
  {{Burovski}}}, \bibinfo {author} {\bibfnamefont {P.}~\bibnamefont
  {{Peterson}}}, \bibinfo {author} {\bibfnamefont {W.}~\bibnamefont
  {{Weckesser}}}, \bibinfo {author} {\bibfnamefont {J.}~\bibnamefont
  {{Bright}}}, \bibinfo {author} {\bibfnamefont {S.~J.}\ \bibnamefont {{van der
  Walt}}}, \bibinfo {author} {\bibfnamefont {M.}~\bibnamefont {{Brett}}},
  \bibinfo {author} {\bibfnamefont {J.}~\bibnamefont {{Wilson}}}, \bibinfo
  {author} {\bibfnamefont {K.}~\bibnamefont {{Jarrod Millman}}}, \bibinfo
  {author} {\bibfnamefont {N.}~\bibnamefont {{Mayorov}}}, \bibinfo {author}
  {\bibfnamefont {A.~R.~J.}\ \bibnamefont {{Nelson}}}, \bibinfo {author}
  {\bibfnamefont {E.}~\bibnamefont {{Jones}}}, \bibinfo {author} {\bibfnamefont
  {R.}~\bibnamefont {{Kern}}}, \bibinfo {author} {\bibfnamefont
  {E.}~\bibnamefont {{Larson}}}, \bibinfo {author} {\bibfnamefont
  {C.}~\bibnamefont {{Carey}}}, \bibinfo {author} {\bibfnamefont
  {{\.I}.}~\bibnamefont {{Polat}}}, \bibinfo {author} {\bibfnamefont
  {Y.}~\bibnamefont {{Feng}}}, \bibinfo {author} {\bibfnamefont {E.~W.}\
  \bibnamefont {{Moore}}}, \bibinfo {author} {\bibfnamefont {J.}~\bibnamefont
  {{Vand erPlas}}}, \bibinfo {author} {\bibfnamefont {D.}~\bibnamefont
  {{Laxalde}}}, \bibinfo {author} {\bibfnamefont {J.}~\bibnamefont
  {{Perktold}}}, \bibinfo {author} {\bibfnamefont {R.}~\bibnamefont
  {{Cimrman}}}, \bibinfo {author} {\bibfnamefont {I.}~\bibnamefont
  {{Henriksen}}}, \bibinfo {author} {\bibfnamefont {E.~A.}\ \bibnamefont
  {{Quintero}}}, \bibinfo {author} {\bibfnamefont {C.~R.}\ \bibnamefont
  {{Harris}}}, \bibinfo {author} {\bibfnamefont {A.~M.}\ \bibnamefont
  {{Archibald}}}, \bibinfo {author} {\bibfnamefont {A.~H.}\ \bibnamefont
  {{Ribeiro}}}, \bibinfo {author} {\bibfnamefont {F.}~\bibnamefont
  {{Pedregosa}}}, \bibinfo {author} {\bibfnamefont {P.}~\bibnamefont {{van
  Mulbregt}}}, \ and\ \bibinfo {author} {\bibfnamefont {S.~.~.}\ \bibnamefont
  {{Contributors}}},\ }\href {\doibase
  https://doi.org/10.1038/s41592-019-0686-2} {\bibfield  {journal} {\bibinfo
  {journal} {Nat. Methods}\ }\textbf {\bibinfo {volume} {17}},\ \bibinfo
  {pages} {261} (\bibinfo {year} {2020})}\BibitemShut {NoStop}%
\bibitem [{\citenamefont {Hindmarsh}(1992)}]{osti_145724}%
  \BibitemOpen
  \bibfield  {author} {\bibinfo {author} {\bibfnamefont {A.}~\bibnamefont
  {Hindmarsh}},\ }\href {https://www.osti.gov/biblio/145724} {\enquote
  {\bibinfo {title} {Odepack. a collection of ode system solvers},}\ }\bibinfo
  {type} {Tech. Rep.}\ (\bibinfo  {institution} {Lawrence Livermore National
  Lab., CA (United States)},\ \bibinfo {year} {1992})\BibitemShut {NoStop}%
\bibitem [{\citenamefont {Byrne}\ and\ \citenamefont
  {Hindmarsh}(1975)}]{byrne1975polyalgorithm}%
  \BibitemOpen
  \bibfield  {author} {\bibinfo {author} {\bibfnamefont {G.~D.}\ \bibnamefont
  {Byrne}}\ and\ \bibinfo {author} {\bibfnamefont {A.~C.}\ \bibnamefont
  {Hindmarsh}},\ }\href {\doibase 10.1145/355626.355636} {\bibfield  {journal}
  {\bibinfo  {journal} {ACM Transactions on Mathematical Software (TOMS)}\
  }\textbf {\bibinfo {volume} {1}},\ \bibinfo {pages} {71} (\bibinfo {year}
  {1975})}\BibitemShut {NoStop}%
\end{thebibliography}%

\end{document}